\documentclass[aip,numerical,reprint,amssymb,amsmath,]{revtex4-1}

\usepackage{graphicx}
\usepackage{subfigure}
\usepackage{balance} 
\usepackage{epsfig}
\usepackage{docs}
\usepackage{bm}
\usepackage[colorlinks=true,linkcolor=blue]{hyperref}
\begin{document}
	
\title{Short time dynamics of water coalescence on a flat water pool}
	
\author{Su Jin \surname{Lim}}
\thanks{These authors contributed equally to this article.}
\affiliation{Soft Matter Physics Laboratory, School of Advanced Materials Science and Engineering, SKKU Advanced Institute of Nanotechnology (SAINT), Sungkyunkwan University, Suwon 16419, Korea}
\author{Bopil \surname{Gim}}
\thanks{These authors contributed equally to this article.}
\affiliation{Department of Bio and Brain Engineering, Korea Advanced Institute of Science and Technology (KAIST), Daejeon 305-701, Korea}
\author{Kamel \surname{Fezzaa}}
\affiliation{X-ray Science Division, Advanced Photon Source, Argonne National Laboratory, 9700 South Cass Avenue, Argonne, Illinois 60439, USA}
\author{Byung Mook \surname{Weon}}
\email{B.M.W. (bmweon@skku.edu)}
\affiliation{Soft Matter Physics Laboratory, School of Advanced Materials Science and Engineering, SKKU Advanced Institute of Nanotechnology (SAINT), Sungkyunkwan University, Suwon 16419, Korea}
	
\date{\today}

\begin{abstract}
Coalescence is an important hydrodynamic event that frequently takes place in nature as well as in industry. Here we provide an experimental study on short time dynamics of water coalescence, particularly when a water droplet comes in contact with a flat water surface, by utilizing high-resolution high-penetration ultrafast X-ray microscopy. Our results demonstrate a possibility that an extreme curvature difference between a drop and a flat surface can significantly modify the hydrodynamics of water coalescence, which is unexpected in the existing theory. We suggest a plausible explanation for why coalescence can be modified by an extreme curvature difference.
\end{abstract}

\maketitle

Coalescence between drops usually takes place to minimize the surface energy. Symmetric coalescence between equal-sized drops \cite{Eggers,Aarts,Case, Paulsen14} has been extensively studied because of its simplicity and importance in natural and industrial situations. Early-time growth of the liquid bridge that accompanies coalescence driven by Laplace pressure has long been a central topic in fluid dynamics with the aim of understanding the relevant coalescence mechanisms. The liquid-bridge growth has been interpreted theoretically throughout integral description in Stokes flow\cite{Eggers} and experimentally with high-speed optical imaging\cite{Aarts} and electrical methods \cite{Case, Paulsen14}. However, asymmetric coalescence between different-sized drops\cite{Thoroddsen05,Thoroddsen07,Zhang09,Hernandez-Sanchez} has been poorly investigated so far, although it is essential to understand various processes such as partial coalescence\cite{Blanchette06,Chen,Blanchette09}, bubble bursting \cite{Zhang08,Lee11} and spreading \cite{Biance,Winkels,Eddi13a,Bonn}. A few recent reports argued an impact of curvature symmetry on the hydrodynamics of water coalescence: curvature asymmetry would affect coalescence\cite{Thoroddsen05,Thoroddsen07} and thus asymmetric coalescence would be different from symmetric coalescence\cite{Zhang09,Hernandez-Sanchez}. More evidence is still required to confirm the effect of curvature asymmetry on coalescence hydrodynamics. Direct visualizations are essential to explain what role curvature asymmetry would play in the coalescence hydrodynamics of water. There is the experimental difficulty in revealing the initial behavior of extremely asymmetric coalescence using traditional imaging methods\cite{Lee11}. Notably, significant progress has been made in X-ray microscopy for direct visualization of rapid hydrodynamic behaviors\cite{Lee11,Fezzaa,Wang,Weon12,Kim}. This approach enables us to investigate the short-time behavior of water coalescence particularly for a water drop on a flat water pool, which is an extreme case of curvature asymmetry. 

In this work, we study experimentally the effect of an extreme curvature difference on water coalescence by utilizing X-ray microscopy. Our results demonstrate the possibility that an extreme curvature difference between drops can play an important role in coalescence hydrodynamics. For extreme asymmetric coalescence, the contact bridge grows rapidly along the water pool surface, which makes a significant difference among extreme asymmetric coalescence, symmetric coalescence, and complete spreading of water. This finding sheds more light on the fundamental understanding of the coalescence hydrodynamics of water.

\begin{figure}
	\centering
	\includegraphics[width=0.44\textwidth]{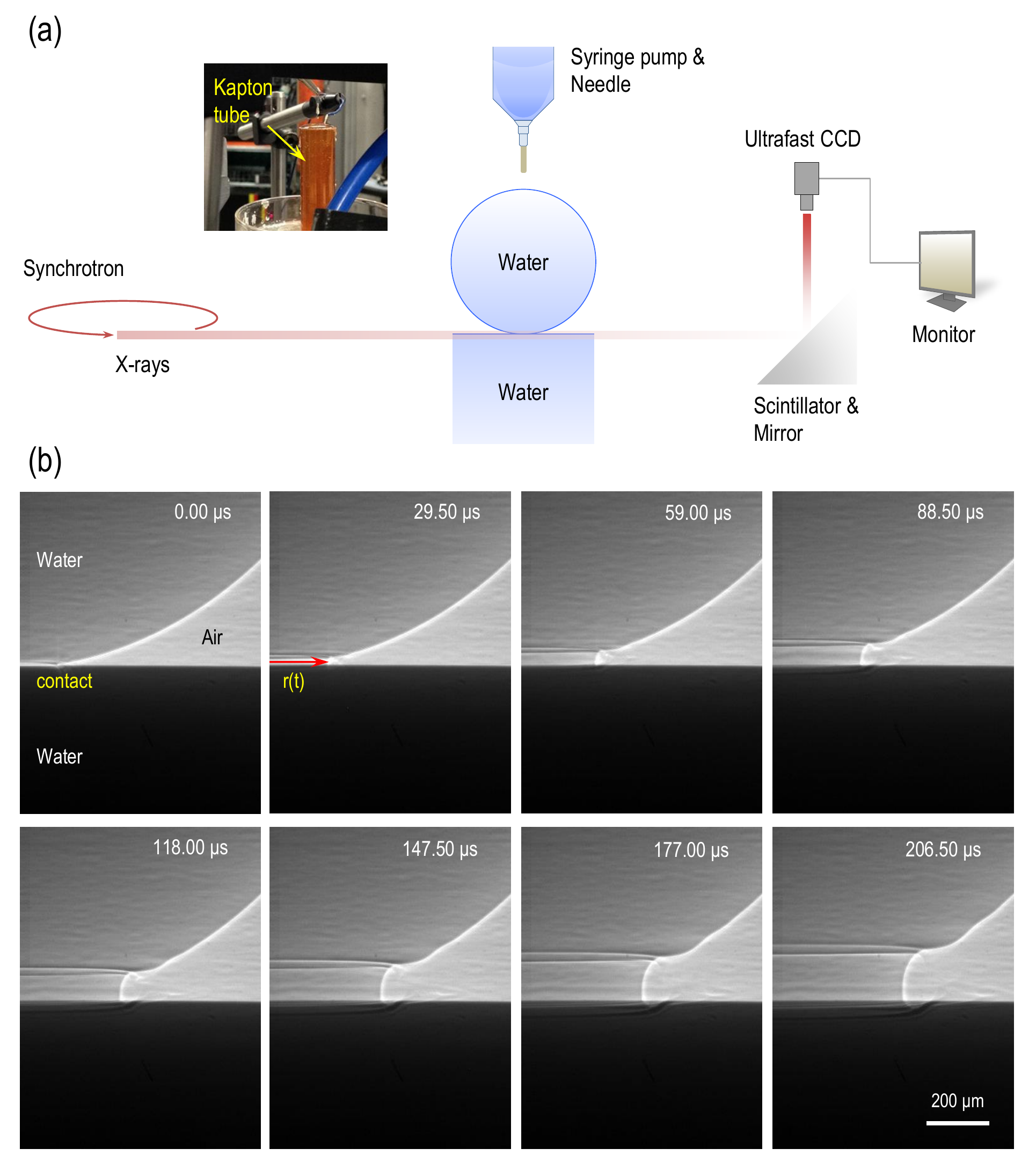}
	\caption{\label{fig:1}Experiments of coalescence with X-ray microscopy. (a) Experimental setup that synchronizes X-ray microscopy and ultrafast CCD camera. Inset image presents Kapton tube with a depth of 50 mm that is transparent to X-rays and can be used for a water pool. (b) An illustration of X-ray sequential images for water with a radius $R \approx$ 1.366 mm on a water pool, surrounded by air (Supplementary Movie 1). The contact bridge radius $r(t)$ grows with time as marked by the arrow after contact.}
\end{figure}

\begin{figure}
	\includegraphics[width=8cm]{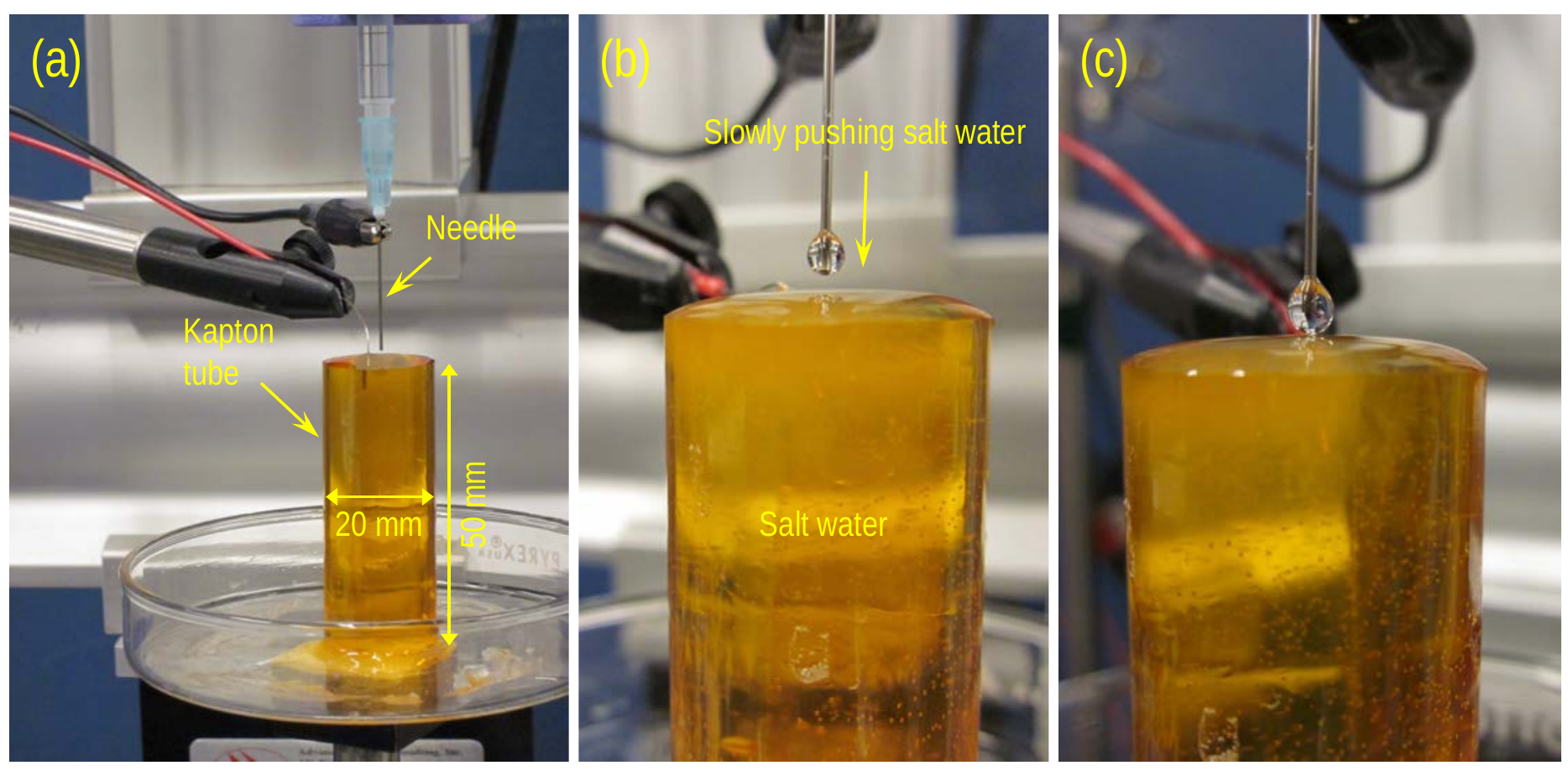}
	\caption{\label{fig:2}When a drop met a flat pool. (a) The water pool was prepared in a cylinder made of Kapton with a diameter of 20 mm and a depth of 50 mm. (b-c) By slowly pushing water from a 26-G syringe needle (outer diameter $\sim$ 0.46 mm and inner diameter $\sim$ 0.26 mm), the water drop gently met the flat water pool. Here the top center of the water pool was almost flat.}
\end{figure}

A series of X-ray imaging experiments were performed at 32-ID-B in the Advanced Photon Source of the Argonne National Laboratory with the experimental setup as schematically illustrated in Fig.~\ref{fig:1}. We modified the setup of Lee {\it et al.}\cite{Lee11} or Wang {\it et al.}\cite{Wang} and used an intense white (full energy spectrum) X-ray beam with a peak irradiance of $\sim$ 10$^{14}$ ph s$^{-1}$mm$^{-2}$ per 0.1 \% bandwidth (bw)\cite{Lee11,Lee15}. The high brightness yielded spatial resolution ($\sim$ 2 $\mu$m per pixel) within a short acquisition time ($\sim$ 14.7 $\mu$s per frame). After passing through the water drop on the water pool (in a Kapton tube), the transmitted X-rays were converted by a scintillator to visible lights, which were then reflected by a mirror and magnified by an objective lens (Mitutoyo M Plan Apo 5, NA $=$ 0.14). After magnification, the image on the scintillator was captured by a CMOS camera (1,024 $\times$ 1,024 pixels; Photron SA 1.1, Photron) that was synchronized with the fast rotary stage and a fast shutter. For synchronization, NaCl was dissolved in pure water to be saturated salt water with 26 \% by mass, to become electrically conductive, as suggested in other studies \cite{Case,Paulsen14}. The whole imaging system was carefully aligned to gravity by using a digital inclinometer with 0.001$^{\circ}$ accuracy. During our experiments, temperature was 23$^{\circ}$C and X-ray heating was negligible, as previously examined at the same conditions\cite{Weon11}. The water pool was prepared in a cylinder made of Kapton with a diameter of 20 mm and a depth of 50 mm, as illustrated in Fig.~\ref{fig:2}. By slowly pushing salt water from a 26-G syringe needle (outer diameter $\sim$ 0.46 mm and inner diameter $\sim$ 0.26 mm), the water drop met the flat water pool. Here the top center of the water pool was almost flat.

\begin{figure}
	\includegraphics[width=8cm]{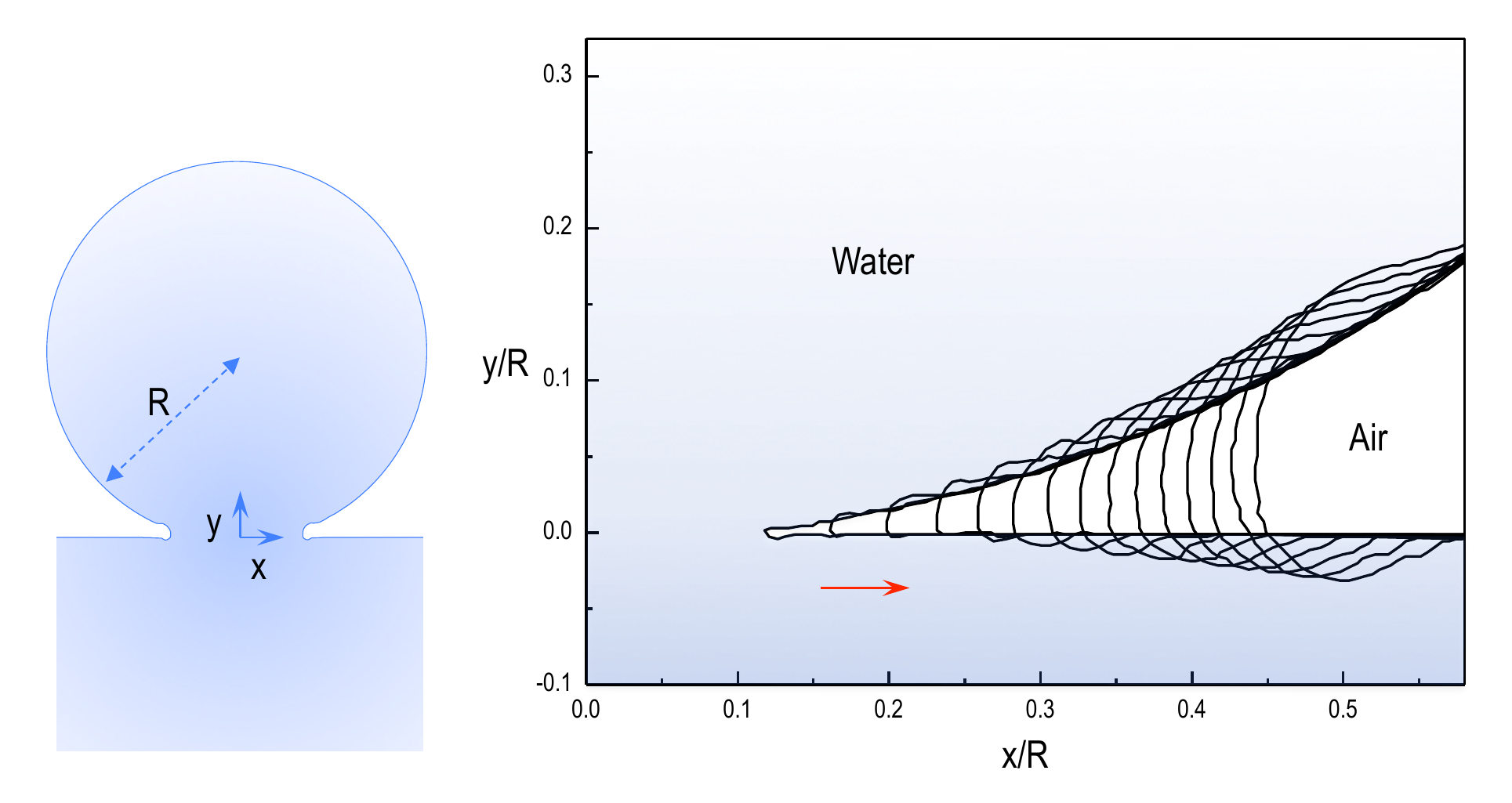}
	\caption{\label{fig:3}Time evolution of water surface profiles. (Left) When a water drop with radius $R$ meets a water pool at a contact point, from which vertical ($y$) and horizontal ($x$) axes originate. (Right) Precise tracking of water surface profiles for upper water, bridge water, and base water: here, profiles are taken from Fig.~\ref{fig:1}(b) (Supplementary Movie 1) with an interval of 29.5 $\mu$s. The arrow indicates the direction of contact bridge growth.}
\end{figure}

Thanks to X-ray microscopy (Fig.~\ref{fig:1}(a)), we directly investigated the short-time behavior of water coalescence for a water drop on a thick water pool that is opaque under optical microscopy. High-resolution and high-penetration X-ray microscopy\cite{Lee11,Fezzaa,Wang} allows for precise tracking of the water surface evolution (Fig.~\ref{fig:1}(b)). X-ray microscopy, which combines X-ray phase-contrast imaging with an ultrafast camera is an elegant solution for identifying the surface profile of water at early times (within 2 ms) with high spatial ($\sim$ 2 $\mu$m per pixel) and temporal (500 ns exposure time and 68 kHz frame rate) resolutions to visualize how a water drop comes in contact with a flat water pool with a relatively slow approaching velocity $U_{app} \approx$ 5.210 $\times$ 10$^{-3}$ m/s (taken from Supplementary Movie 2), which might induce collision effects \cite{Case,Paulsen,Duchemin}. We calibrated the contact bridge radius $r(t)$ by subtracting the original radius $r_{i}(t)$ from the collision-mediated radius $r_{j}(t) = \sqrt{U_{app}Rt}$ for a given drop with a radius $R$ with time $t$\cite{Duchemin}. We used $R \approx$ 1.366 or 1.540 mm-sized drops on average. X-ray imaging is capable of differentiating the profile of a water drop from a contact bridge with radius $r$ beneath a flat water surface, as demonstrated in Fig.~\ref{fig:3}. The asymmetric water profiles appear between an upper drop and a base pool.

The quantitative analyses in Figs.~\ref{fig:1} and ~\ref{fig:3} lead to the conclusion that the bridge radius grows rapidly at a speed $v \sim2.240$ m/s at very early stages. This speed indicates the inertial coalescence that takes place at large Reynolds numbers, Re $= \frac{\rho wv}{\eta} > 1$ (where the bridge height $w \geq$ 0.001 mm, fluid density $\rho$ = 1197.2 kg/m$^{3}$, and fluid viscosity $\eta$ = 1.662 mPa $\cdot$ s for salt water\cite{Paulsen}), in our experiments, where capillary and inertial forces are dominantly competing, but viscous forces are negligible. Consequently we expect a scaling as:
\begin{equation}
r \sim \Bigl(\frac{\sigma R}{\rho}\Bigr)^{1/4} t^{1/2},
\label{eq:1}
\end{equation}
where the prefactor is associated with the drop radius and hydrodynamic parameters of surface tension $\sigma$, fluid density, and fluid viscosity. The most important fact is that this scaling law, the $\sqrt t$ dependence of $r$, is widely accepted to explain both the hydrodynamics of drop coalescence \cite{Eggers,Aarts} and the short-time behavior of spreading water drops surrounded by air \cite{Biance, Winkels,Eddi13a,Bonn}.

\begin{figure}
	\includegraphics[width=7cm]{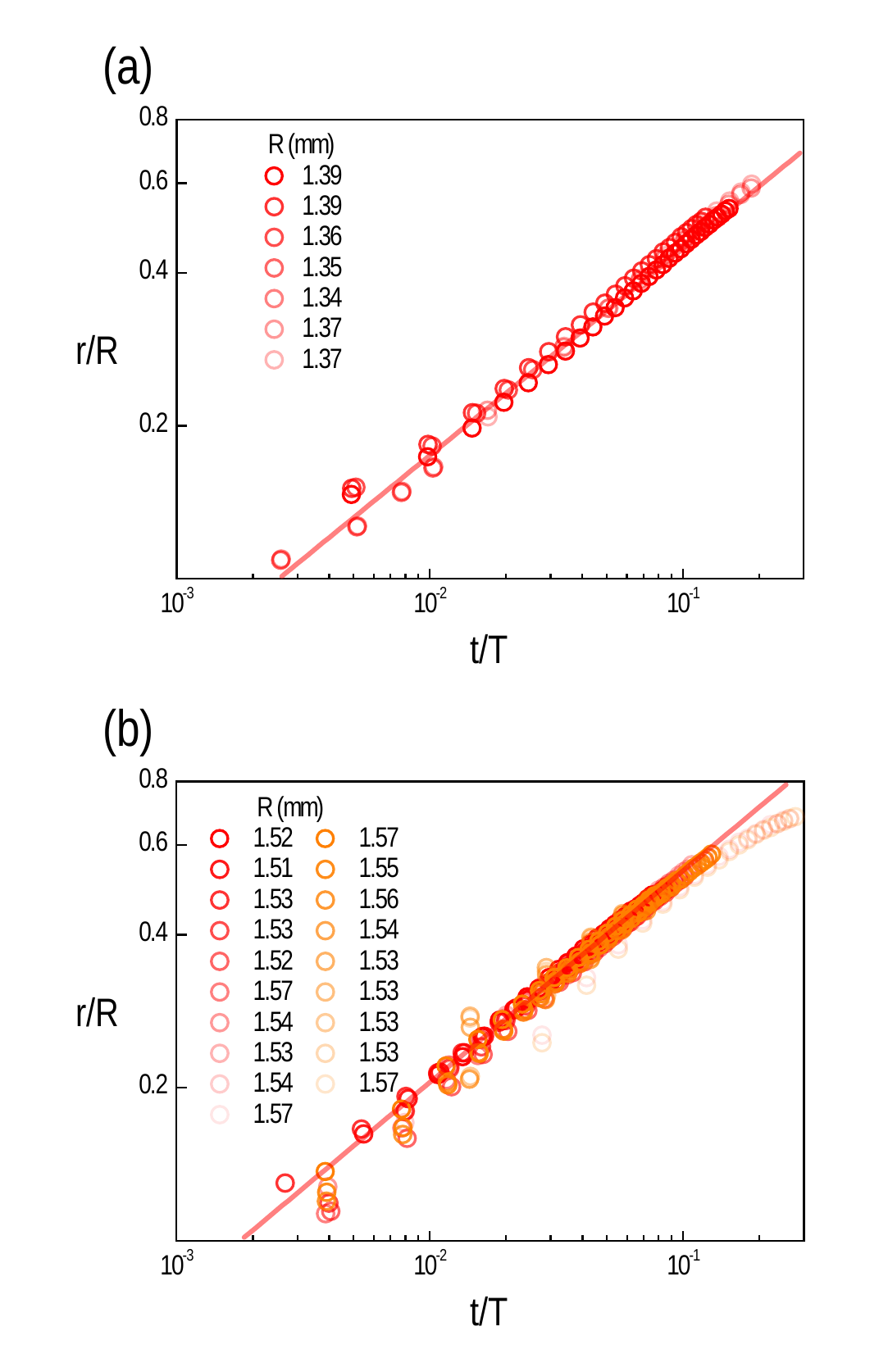}
	\caption{\label{fig:4}Experimental results of extreme asymmetric coalescence hydrodynamics. The equation $(r/R) \sim (t/T)^{\xi}$ is tested with two distinct experimental datasets of (a) $R \sim$ 1.366 mm for 7 cases and (b) $R \sim$ 1.540 mm for 19 cases. The slopes $\xi$ are measured as (a) $0.418 \pm 0.004$ and (b) $0.414 \pm 0.002$, as depicted by the solid lines.}
\end{figure}

In our observations, we found that the $\sqrt t$ dependence of $r$ is invalid for a base water pool. From now, we rewrite Eq.~\ref{eq:1} for general formula to cover even asymmetric coalescence: 
\begin{equation}
(r/R) \sim (t/T)^{\xi}
\label{eq:2}
\end{equation}
with $T=\sqrt{\rho R^3/ \sigma}$ called inertial time \cite{Aarts}, where $\sigma = 88.55$ mN/m for salt water \cite{Paulsen}.  The simple power law of Eq.~\ref{eq:2} was valid for the almost whole range of $T$. Here, $\xi$ is the growth exponent and it has a value of $1/2$ for symmetric coalescence or spreading of water as in Eq.~\ref{eq:1}. In Figs.~\ref{fig:4}(a) and~\ref{fig:4}(b) of the extremely asymmetric coalescence dynamics, $\xi = 0.418\pm0.004$ and $0.414\pm0.002$ were respectively measured from our observations. These values ($\approx 2/5$) are slightly different from $1/2$ which has been broadly accepted in symmetric coalescence or spreading.

Here, we discuss why symmetric coalescence between equal-sized water drops and spreading of a water drop on a completely wettable solid surface show a similar initial behavior. Both cases exhibit the $\sqrt t$ dependence of $r$ ($\xi \approx 1/2$). Spreading of a water drop on a perfectly wettable solid surface is much alike to a half part of symmetric coalescence of water drops. This case is valid only if the bottom solid does not disturb the short-time spreading dynamics.

A general explanation can be given on the basis of recent works\cite{Duchemin,Paulsen}. For symmetric coalescence, a fluid bridge of radius $r$ has height $w = r^{2}/R$ \cite{Duchemin}. As the bridge expands radially, fluid fills the neighboring region. The speed of this flow $U$ is proportional to $dr/dt$. The volume sandwiched between the drops out to a radius $r$ is given by $V = \frac{\pi}{2R}r^{4}$ and evolves as $dV/dt = \frac{2\pi}{R}r^{3}dr/dt$ \cite{Paulsen}. This fluid is supplied through two annular regions of radius $r$ and width $w$, above and below the gap, of total area $A = \frac{4\pi}{R}r^{3}$ \cite{Paulsen}. Setting the volume flow rate $AU$ equal to $dV/dt$, we get $U \approx \frac{1}{2}dr/dt$ \cite{Paulsen}. For complete spreading, only the upper side exists, so that $w = \frac{1}{2}r^{2}/R$, $V = \frac{\pi}{4R}r^{4}$, and $A = \frac{2\pi}{R}r^{3}$. Setting $AU = dV/dt$, we derive $U \approx \frac{1}{2}dr/dt$ for complete spreading. Therefore, both cases have the identical flow speed $U$ that drives the identical bridge growth, that is, the $\sqrt t$ dependence of $r$ ($\xi \approx 1/2$). 

\begin{figure}
	\includegraphics[width=8cm]{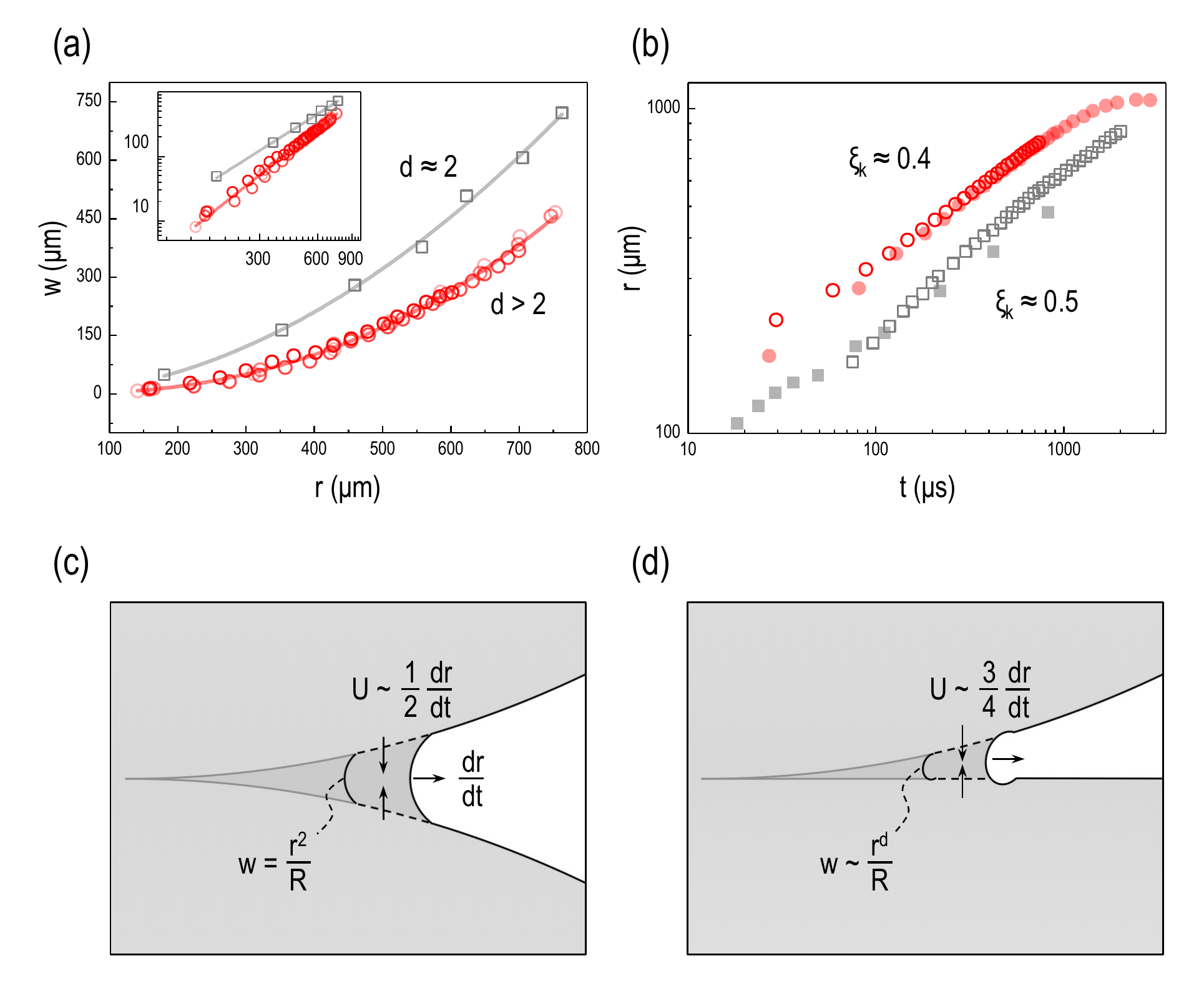}
	\caption{\label{fig:5}Comparisons of our and other observations. (a) The bridge height $w$ measured from \cite{Aarts} for symmetric coalescence (squares) and fitted by $w \sim r^{d}$ (solid lines) shows $d =  1.904 \pm 0.066$, while the $w$ values taken from 4 datasets in Fig.~\ref{fig:4}(a) for asymmetric coalescence (circles) show $d = 2.389 \pm 0.033$. (b) The bridge radius $r(t)$ (open circles) is consistent with \cite{Thoroddsen07} (closed circles) for asymmetric coalescence ($\xi_{k} \approx 0.4$) and is placed higher than for symmetric coalescence ($\xi_{k} \approx 0.5$, open and closed squares) \cite{Thoroddsen07,Fezzaa}. (c-d) Schematic illustrations for symmetric coalescence\cite{Paulsen} in (c) and asymmetric coalescence in (d), suggesting a rapid bridge growth for asymmetric coalescence.}
\end{figure}

For asymmetric coalescence, the bridge height $w$ is not exactly equal to $r^{2}/R$ but proportional to $r^{d}/R^{d-1}$. Our observations in Fig.~\ref{fig:5}(a) show that $d > 2$ for extreme asymmetric coalescence (circles, data taken from Fig.~\ref{fig:4}(a)), suggesting that $w \approx r^{d}/R^{d-1}$, compared with $d \approx 2$ for symmetric coalescence (squares, data taken from Ref.~\onlinecite{Aarts}). The volume between the drop and the flat surface is given as $V \sim \frac{\pi}{2R^{d-1}}r^{d+2}$ and the area corresponding to the volume flow is approximated as $A \sim \frac{3\pi}{R^{d-1}}r^{d+1}$ where the lower area is smaller than the upper area because of the extreme asymmetric geometry. Setting $AU = dV/dt$, we get $U \approx \frac{(d+2)}{6}dr/dt$. Approximating $d \approx 2.5$ based on Fig.~\ref{fig:5}(a), we finally get $U \approx \frac{3}{4}dr/dt$. This speed is slightly faster than the speed for the volume supply in symmetric coalescence. This result means that the rapid volume supply by $U$ would drive the fast bridge growth in asymmetric coalescence.

Evidently, the $r/R$ versus $t/T$ curves for extreme asymmetric cases are placed higher than for symmetric cases in Fig.~\ref{fig:5}(b). The $r/R$ from our results in Fig.~\ref{fig:5}(b) is consistent with that of other observations for extreme asymmetric coalescence ($\xi \approx 0.4$) \cite{Thoroddsen07} and is higher than for symmetric coalescence ($\xi \approx 0.5$)\cite{Thoroddsen07,Fezzaa}. As for the geometry effect, the bridge growth is illustrated in Fig.~\ref{fig:5}(c) for symmetric coalescence and in Fig.~\ref{fig:5}(d) for extreme asymmetric coalescence. The contact bridge grows rapidly along the water pool surface, which makes a significant difference between symmetric and extreme asymmetric coalescence.

The deviation of $\xi$ values from 0.5 would be confirmed from our results, which is clear from additional analyses for asymmetric coalescence data in Fig.~\ref{fig:5}(b) and for each datasets in Fig.~\ref{fig:4}(b) with $\xi$ (variable, Supplementary Table 1) and $\xi = 0.5$ (fixed, Supplementary Table 2). Here, $r \approx C_{a} t^{0.4}$ and $r \approx C_{s} t^{0.5}$ are equivalent to $(dr/dt)_{a} \approx 0.4 C_{a} t^{-0.6}$ and $(dr/dt)_{s} \approx 0.5 C_{s} t^{-0.5}$ for asymmetric (denoted by $a$) and symmetric (denoted by $s$) cases. Consequently, $r(t)_{a} > r(t)_{s}$  (Fig.~\ref{fig:5}(b)) and $(dr/dt)_{a} \gtrsim (dr/dt)_{s}$, which is possible because of $C_{a} > C_{s}$. The differences in the prefactor and the exponent should be further studied in detail. With respect to the finite-size effect \cite{Gross}, the range of experiments needs to be extended to confirm the $\xi$ values.

Finally, we discuss the other effects such as vortex ring and outer surrounding fluid. The bridge height would be relevant to the capillary waves along the bottom water pool. The coalescence of a drop with a pool is known to produce a vortex ring \cite{Dooley,Hamlin} but absent at very early stages (within 1 ms) of coalescence. The outer fluid (air in our case, whereas oil in in Ref.~\onlinecite{Aryafar}) would be important for the bridge growth in asymmetric coalescence. Further study is required with X-ray microscopy by widely varying two curvatures between drops for different fluids.

To conclude, the short-time dynamics of water coalescence for a water drop on a flat water pool is revealed with X-ray imaging. This work shows the effect of an extreme curvature difference on coalescence hydrodynamics; extreme curvature asymmetry between drops can slightly modify the hydrodynamics of water coalescence, by which the analogy between coalescence and spreading can be broken. We believe that an extreme curvature difference would play an essential role in the hydrodynamic behaviors of water or other liquids in a variety of natural and industrial processes.

See supplementary material for representative movies (Movies 1 and 2) taken by X-ray microscopy and for the detailed results (Tables 1 and 2; Figures S1 and S2).
	
We would like to thank the APCTP for its hospitality. Use of the Advanced Photon Source, an Office of the Science User Facility operated by the US Department of Energy (DOE) Office of Science by Argonne National Laboratory, was supported by the US DOE under contract no. DE-AC02-06CH11357.

\end{document}